\documentclass{article}
\usepackage{spconf,amsmath,graphicx,hyperref}

\usepackage{booktabs}
\usepackage{xcolor}
\usepackage{siunitx}
\usepackage{multirow}
\title{SEA-Spoof: Bridging the Gap in Multilingual Audio Deepfake Detection for South-East Asia}
\name{\begin{tabular}{@{}c@{}}
Jinyang Wu$^{1}$, 
Nana Hou$^{2}$,
Zihan Pan$^{1*}\thanks{*Corresponding author}$,
Qiquan Zhang$^{3}$,
Sailor Hardik Bhupendra$^{1}$,
Soumik Mondal$^{1}$
\end{tabular}}

\address{$^1$Institute for Infocomm Research (I2R), A*STAR, Singapore,\\
$^2$Nanyang Technological University,
$^3$The University of New South Wales,\\
\{wu\_jinyang, panz\}@a-star.edu.sg
}
\date{September 24, 2025}
\begin{document}
\ninept
\maketitle
%


\begin{abstract}
The rapid growth of the digital economy in South-East Asia (SEA) has amplified the risks of audio deepfakes, yet current datasets cover SEA languages only sparsely, leaving models poorly equipped to handle this critical region. This omission is critical: detection models trained on high-resource languages collapse when applied to SEA, due to mismatches in synthesis quality, language-specific characteristics, and data scarcity. To close this gap, we present \textbf{SEA-Spoof}, the first large-scale Audio Deepfake Detection (ADD) dataset especially for SEA languages. SEA-Spoof spans 300+ hours of paired real and spoof speech across Tamil, Hindi, Thai, Indonesian, Malay, and Vietnamese. Spoof samples are generated from a diverse mix of state-of-the-art open-source and commercial systems, capturing wide variability in style and fidelity.
Benchmarking state-of-the-art detection models reveals severe cross-lingual degradation, but fine-tuning on SEA-Spoof dramatically restores performance across languages and synthesis sources. These results highlight the urgent need for SEA-focused research and establish SEA-Spoof as a foundation for developing robust, cross-lingual, and fraud-resilient detection systems.
\end{abstract}

\section{Introduction}
\label{sec:intro}

Generative modeling approaches have recently demonstrated significant advancements in text-to-speech (TTS) and voice cloning (VC), resulting in increasingly realistic spoofed speech. They advance accessibility, entertainment, and personalized services, however, facilitate the creation of audio deepfakes, thereby posing urgent security concerns. As a result, detecting manipulated speech has emerged as a central challenge for the speech community.  


The progress in speech fake detection has been driven largely by benchmark datasets. Specifically, the ASVspoof has provided increasingly large-scale and challenging corpora. ASVspoof-2019~\cite{asv2019} introduced the widely adopted logical access (LA) and physical access (PA) tracks, covering diverse TTS, VC, and replay attacks. ASVspoof-2021~\cite{asv2021} further expanded the benchmark with a new deepfake (DF) track and emphasized cross-database generalization, incorporating more advanced neural TTS/VC systems. Most recently, ASVspoof5~\cite{asv5} scaled up to hundreds of thousands of fake samples generated with a wide spectrum of modern architectures (e.g., GlowTTS, VITS, XTTS, YourTTS, DiffVC) and even adversarial perturbations, while introducing combined countermeasure and spoof-robust ASV evaluation protocols. The spoofing systems in ASVspoof5 span autoregressive, non-autoregressive, diffusion-based, and neural vocoder architectures, thereby ensuring broad coverage of contemporary speech synthesis techniques. These benchmarks established standard protocols and advanced the field significantly, but still mainly focus on English and a few high-resource languages.  A number of datasets have been recently released, such as DFADD~\cite{dfadd2024}, CodecFake~\cite{codecfake2024}, and SpoofCeleb~\cite{jung2024spoofceleb}, which broaden the coverage of attack types and codecs but remain limited in language diversity. In-The-Wild~\cite{jung2024textwild} offers realistic fake samples, yet remains primarily focused on English.



 In recent years, multilingual speech deepfake detection benchmarks have emerged, but their scope remains limited. M-AILABS~\cite{mailabs} provides multilingual data, yet with small scale and narrow linguistic coverage. MLAAD~\cite{mlaad2024} extends to more languages, but its size is modest and its representation incomplete. SpeechFake~\cite{huang2025speechfakelargescalemultilingualspeech} represents the largest multilingual dataset to date, but its emphasis is on generation methods rather than regional balance, and it does not specifically address low-resource or South-East Asian (SEA) languages.


This gap is particularly acute in South-East Asia (SEA), where detection models trained on high-resource languages often fail due to mismatches in synthesis quality, language-specific characteristics, and limited training resources \cite{wu2015spoofing}. SEA languages also pose unique challenges: tonal languages such as Thai and Vietnamese differ fundamentally from non-tonal ones like Hindi, Tamil, Malay, and Indonesian, creating diverse prosodic and phonological artifacts. Together, these factors underscore the need for a dedicated dataset that captures both linguistic and system-level diversity rather than assuming transferability from English or Chinese.  

To close this gap, in this paper we present \textbf{SEA-Spoof}, the first large-scale dataset for speech deepfake detection across six SEA languages, i.e., Tamil, Hindi, Thai, Indonesian, Malay, and Vietnamese. SEA-Spoof pairs real recordings with spoof speech generated from 10 state-of-the-art (SoTA) open-source and 4 commercial systems, spanning both TTS and VC, thereby ensuring broad coverage of generation paradigms and controlled real–fake comparisons.  


In summary, our main contributions are as follows:  
\begin{itemize}
    \item To the best of our knowledge, we release the first balanced, large-scale dataset for fake speech detection in six SEA languages, covering both TTS and VC with diverse open-source and commercial systems.  
    \item We benchmark state-of-the-art models, revealing severe mismatch performance gaps and demonstrating how SEA-Spoof restores robustness when used for fine-tuning.  
\end{itemize}

By filling this critical regional and linguistic gap, SEA-Spoof establishes a foundation for building robust, cross-lingual, and fraud-resilient audio deepfake detection systems, and calls attention to multilingual security in speech AI.

\begin{figure*}[!t]
    \centering
    \includegraphics[scale=0.8, trim=80 255 20 190, clip]
     {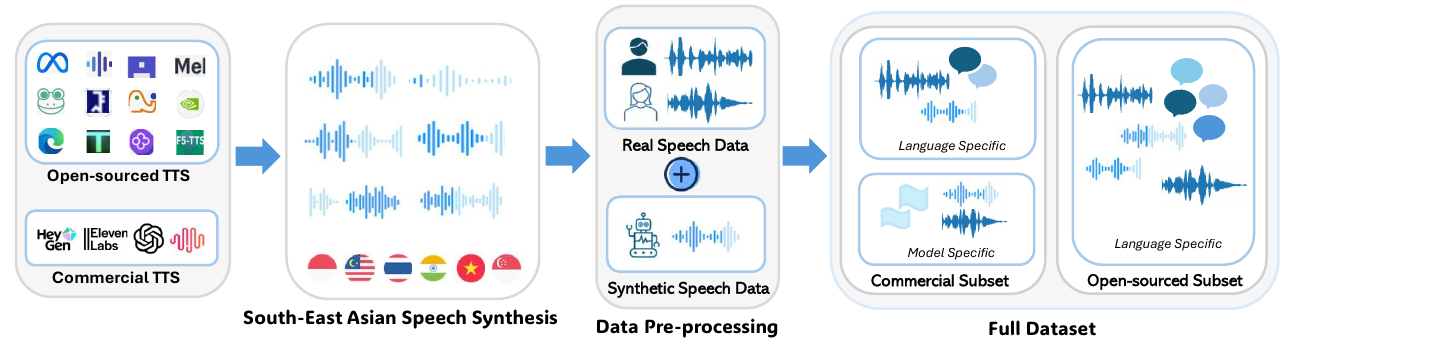}
    \caption{Pipeline of the SEA-Spoof dataset, combining speech synthesized from open-source and commercial TTS with real recordings, and organized into language- and system-specific subsets for evaluation.}
    \label{fig:placeholder}
\end{figure*}

\section{Dataset Construction}
\label{sec:method}

\subsection{Language Selection}
We include six widely spoken languages in South-East Asia: Tamil, Hindi, Thai, Indonesian, Malay, and Vietnamese. Three considerations guide the selection. 
(i) \textit{Population coverage}: these languages are spoken by hundreds of millions across SEA and neighboring regions, including significant diaspora communities (e.g., Tamil and Hindi in Singapore and Malaysia).  
(ii) \textit{Linguistic diversity}: the set spans tonal languages (Thai, Vietnamese \cite{vietnamese_lang}) and non-tonal ones (Indonesian \cite{indonesian_lang}, Malay, Tamil, Hindi), which differ substantially in prosody and phonetics, thereby creating diverse synthesis and detection challenges.  
(iii) \textit{Practical relevance}: these languages are increasingly targeted in real-world TTS applications and online fraud scenarios, yet remain largely absent from existing fake speech datasets.  

By combining population significance, linguistic diversity, and practical relevance, our dataset captures both regional importance and technical challenge, providing a critical resource for advancing fake speech detection in SEA.

\begin{table*}[!t]
\centering
\small
\def\arraystretch{1.3}
\setlength{\tabcolsep}{2.5pt}
\setlength{\abovetopsep}{0pt}
\setlength\belowbottomsep{0pt} 
\setlength\aboverulesep{0pt} 
\setlength\belowrulesep{0pt}
\caption{Statistics of the SEA Fake Speech dataset. Open-source models are denoted as A1–A10 and commercial models as C1–C4. Bona fide recordings are drawn from seven open-source corpora and paired with synthetic counterparts to form balanced real–fake data. Durations are reported separately for open-source and commercial sources.}
\label{tab:sea_data_statistics}
\vspace{0.1in}
\resizebox{0.93\textwidth}{!}{%
\begin{tabular}{lccccc}
\toprule[1.1pt]
\multirow{2}{*}{\textbf{Language}} & \multirow{2}{*}{\textbf{Open-sourced Models}} & \multirow{2}{*}{\textbf{Commercial Models}} & \textbf{Training Data (hours)} & \textbf{Validation Data (hours)} & \textbf{Testing Data (hours)} \\
\cline{4-6}
 & & & \multicolumn{3}{c}{Open-sourced / Commercial} \\
\midrule
Malay      & A1, A4, A6, A8         & C1--C4 & 54.4/11.1     & 6.8/1.39   & 6.8/1.39 \\
Indonesian & A1, A4, A6             & C1--C4 & 48/13.36      & 6.0/1.67   & 6.0/1.67 \\
Thai       & A1, A2, A3, A6         & C1--C4 & 44.8/14.0     & 5.6/1.75   & 5.6/1.75 \\
Hindi      & A1, A4, A5, A6, A7, A9 & C1--C4 & 47.2/17.04    & 5.9/2.13   & 5.9/2.13 \\
Tamil      & A1, A5, A6, A7, A9, A10& C1--C4 & 44.8/13.6     & 5.6/1.7    & 5.6/1.7  \\
Vietnamese & A1, A4, A6             & C1--C4 & 40.0/8.0      & 5.0/1.0    & 5.0/1.0  \\
\midrule
Total Spoofed Data & A1-A10 & C1-C4 & 356.3 & 44.54 & 44.54 \\
Total Bona-fide Data & B1-B7 & -- & 354.8 & 44.35 & 44.35 \\

Total Hours & A1-A10, B1-B7 & C1-C4 & 711.1 & 88.89 & 88.89 \\
Total Utterances & A1-A10, B1-B7 & C1-C4 & 439718 & 57245 & 57302 \\
Ratio: Spoofed / Bona-fide & A1-A10, B1-B7 & C1-C4 & 1.004 & 1.004 & 1.004 \\
\toprule[1.3pt]
\end{tabular}
}
\end{table*}

\subsection{Spoof Data}
To capture the diversity of modern speech generation and enhancement \cite{nanaFeature_Representations}, we synthesized audio using 10 open-source and 4 commercial models, spanning text-to-speech (TTS) and voice cloning approaches. These systems encompass autoregressive, non-autoregressive, diffusion-based, and neural vocoder architectures, providing a representative spectrum of current synthesis technologies across SEA languages. Each sample was paired with a corresponding real utterance, ensuring balanced coverage of both open-source and commercial sources. All spoof waveforms were resampled to 16 kHz and stored in FLAC format.

\subsubsection{Open-Source Models}
We employed widely adopted open-source frameworks, including VITS-MMS (A1) \cite{pratap2023vist-mms}, Edge-TTS (A2) \cite{edge-tts}, XTTS-v2 (Coqui) (A3) \cite{xtts_Coqui_TTS_2021}, FastSpeech2 (A4) \cite{ren2022fastspeech2}, Indic-TTS (A5)~\cite{indictts2023}, F5-TTS (A6)~\cite{f5-tts}, Tacotron2 (A7) \cite{shen2018naturalttstacotron2}, MelGAN-FastSpeech2 (A8) \cite{kumar2019melgan}, FastPitch (A9)~\cite{fastpitch}, and Glow-TTS (A10) \cite{kim2020gGlow-TTS}. These systems differ in generation pipeline (autoregressive vs.\ non-autoregressive) and vocoder backends. Although not all natively support SEA languages, we enabled them via fine-tuning on regional corpora (e.g., Indic-TTS for Hindi/Tamil). Among them, VITS-MMS and Edge-TTS provide the broadest multilingual coverage and serve as strong baselines for SEA speech synthesis.  

\subsubsection{Commercial Models}
To reflect real-world black-box scenarios, we integrated four commercial platforms: HeyGen (C1) \cite{heygen}, ElevenLabs (C2) \cite{ElevenLabs}, MiniMax (C3) \cite{MiniMax}, and ChatGPT-4o-mini-TTS (C4) \cite{ChatGPT-4o-mini-TTS}. These systems offer higher perceptual quality, richer speaker diversity, and controllable speaking styles, making them indispensable for evaluating cross-source generalization of detection models.  

\subsubsection{Spoof Data Generation}
\textbf{Text-to-Speech.} We used VITS-MMS (A1) as the backbone for multilingual generation, supplemented with FastSpeech2 (A4), Indic-TTS (A5), F5-TTS (A6), Tacotron2 (A7), MelGANFastSpeech2 (A8), FastPitch (A9), and Glow-TTS (A10). Language-specific Coqui models (A3) were also integrated for Thai and Indonesian. Commercial systems (C1–C4) provided additional variability and high-fidelity speech.  

\noindent \textbf{Voice Cloning.} For cross-speaker synthesis, we employed XTTS-v2 (A3) and Edge-TTS (A2) alongside the four commercial platforms (C1–C4). To enrich variability, we intentionally generated speech in SEA target languages using non-native speakers. This cross-language transfer introduced accented and style-diverse fakes, better reflecting real-world misuse and providing more challenging cases for detection.

\subsection{Bona-fide Data}

To provide real counterparts, we curate Bona-fide recordings from multiple corpora originally designed for TTS and ASR tasks. TTS corpora contribute studio-quality, highly intelligible speech, whereas ASR corpora offer more diverse, noisy, and conversational material. This mix captures both controlled and real-world conditions for robust evaluation.

Our largest source is Mozilla Common Voice~\cite{ardila-etal-2020-common} (B1), which spans all six SEA languages with multi-speaker coverage. For Hindi and Tamil, we additionally use the Indic Speech Corpora \cite{indictts2023} (B2); for Indonesian, we adopt GigaSpeech2 \cite{gigaspeech2} (B3), a large-scale, multi-domain collection; for Malay, we combine the Malay Conversational Speech Corpus\footnote{The dataset is available at \url{https://magichub.com/datasets/malay-conversational-speech-corpus/}} (B4) with a curated Malaysian YouTube Whisper-Large set\footnote{The dataset is available at \url{https://huggingface.co/datasets/mesolitica/}} (B5) to provide diverse accents and conversational styles; for Thai, we include the Thai Dialect Corpus \cite{thaidia23_interspeech} (B6), covering both standard and regional varieties; and for Vietnamese, we use the VIVOS corpus \cite{vivosviet} (B7). Together, these resources provide broad coverage of speakers, domains, and recording conditions, ensuring that SEA-Spoof reflects realistic distributions rather than purely spoof artifacts.

To ensure strict pairing between Bona-fide and spoof speech, we use the transcripts of real utterances as textual prompts when generating the corresponding spoof samples (via the TTS/VC systems described in Section~\ref{sec:method}; A1–A10, C1–C4). Grounding synthesis in real transcripts maximizes comparability across real–fake pairs, enabling controlled evaluations where differences can be attributed directly to synthesis artifacts and yielding rigorous, interpretable benchmarking for detection models.

We will release SEA-Spoof for research purposes.\footnote{The dataset will be avaibale at \url{https://huggingface.co/datasets/Jack-ppkdczgx/SEA-Spoof/}}

\section{Experimental Results}
\label{sec:experiment}

\subsection{Dataset Overview}
The SEA-Spoof dataset provides large-scale, balanced coverage of six South-East Asian languages: Malay, Indonesian, Thai, Hindi, Tamil, and Vietnamese. For each language, synthetic speech is generated by multiple open-source (A1–A10) and commercial (C1–C4) systems, paired with authentic recordings to form real–fake pairs.
\begin{itemize}
    \item Scale: The dataset totals 711 hours of audio, comprising 356.3 hours spoofed and 354.8 hours bona-fide speech, with a spoofed-to-bona-fide ratio of 1:1 across all splits.
    \item Splits: The dataset is partitioned into training (439k utterances), validation (57k utterances), and testing (57k utterances) sets, following an 8:1:1 ratio.
    \item Diversity: Open-source models span autoregressive, non-autoregressive, and diffusion-based architectures, while commercial systems provide black-box, high-quality speech with diverse speakers and styles.
    \item Granularity: Beyond the full dataset, SEA-Spoof also provides language-specific and system-specific subsets (see Sec. 3), enabling fine-grained benchmarking across synthesis conditions.
\end{itemize}  
Table~\ref{tab:sea_data_statistics} summarizes dataset statistics per language, system group, and split. This design ensures that SEA-Spoof captures both linguistic diversity and synthesis variability, offering a rigorous testbed for evaluating cross-lingual and cross-source generalization in audio deepfake detection.

%

\subsection{Baselines}
We benchmarked SEA-Spoof using two widely adopted baseline anti-spoofing models and one state-of-the-art model.
\begin{itemize}
    \item \textbf{AASIST} \cite{jung2021aasist}: a graph attention network designed for replay and spoof spoofing detection trained on ASVspoof 2019 logical access (LA) dataset.  
    \item \textbf{AASIST3} \cite{borodin2024aasist3kanenhancedaasistspeech}: an improved variant incorporating kernel attention and enhanced spectral modeling trained on a combination of multiple datasets: ASVspoof 2019 LA, ASVspoof 5, MLAAD (Multi-Language Audio Anti-Spoofing Dataset).
    \item \textbf{MoLEx} \cite{moeloraantispoof}: a mixture-of-experts model with a LoRA-based fast adaptation module, leverages diverse encoders for robust spoof detection trained on ASVSpoof 5, ASVSpoof 2019, DFADD \cite{dfadd2024}, Fake-or-Real \cite{aptly-fake-or-real-2025} datasets.
\end{itemize} 

For training and evaluation with MoLEx model, we adopt the MoLEx framework with a 12-layer transformer backbone. The classifier consists of a single LSTM layer with hidden size 192 followed by a fully connected layer with output dimension 2. During optimization, only the MoLEx modules, attentive merging block, and classifier are updated, while the pretrained backbone remains frozen. The model is trained with batch size 500 for 15 epochs. MoE-specific settings include 12 experts, 12 MoE layers, expert rank 32, top-k=4, and an extended expert factor of 1. Following \cite{moeloraantispoof}, we use WavLM as both the architecture and self-supervised learning (SSL) feature extractor.

\begin{table}[!t]
\centering
\small
\def\arraystretch{1.11}
\setlength{\tabcolsep}{2.5pt}
\setlength{\abovetopsep}{0pt}
\setlength\belowbottomsep{0pt} 
\setlength\aboverulesep{0pt} 
\setlength\belowrulesep{0pt}
\caption{EER (\%) results of different models on ASVspoof5 and SEA-Fake full datasets.}
\vspace*{0.05in}
\begin{tabular}{lcccc}
\toprule[1.0pt]
\textbf{Dataset} & \textbf{AASIST} & \textbf{AASIST3} & \textbf{MoLEx} & \textbf{MoLEx Fine-tuned} \\
\hline
ASVspoof5      & 35.5\%  & 34\%    & 1.25\%   & 5.6\%  \\
SEA-Fake (a)& 43.32\% & 33.93\% & 43.822\% & 0.2\%  \\
\toprule[1.0pt]
\end{tabular}
\label{tab:experiment1_asvspoof5}
\end{table}

\subsection{Results}
\subsubsection{Mismatch Analysis}

We first evaluated AASIST, AASIST3, and MoLEx on both the ASVspoof5 test set and the SEA-Spoof test set. Results (Table~\ref{tab:experiment1_asvspoof5}) reveal severe cross-domain mismatch: MoLEx achieves strong performance on ASVspoof5 (1.25\% EER) but collapses on SEA-Spoof (43.8\% EER). AASIST and AASIST3 show similar degradation. These results confirm that current benchmarks overlook the unique challenges of SEA languages, leading to poor generalization.  

\subsubsection{Gap-Bridge Analysis}

We then fine-tuned MoLEx on SEA-Spoof. As shown in Table~\ref{tab:experiment1_asvspoof5}, the fine-tuned model (MoLEx\(_{ft}\)) restores near-perfect detection on SEA-Spoof (0.2\% EER). While its performance on ASVspoof5 drops slightly due to catastrophic forgetting, this can be mitigated with mixed-source training, which we leave for future exploration. This demonstrates that SEA-Spoof effectively bridges the generalization gap caused by cross-lingual and cross-source mismatches.  

\subsubsection{Spoof Model Analysis}

To analyze spoofing difficulty, we split SEA-Spoof into subsets: (i) fakes from open-source models and (ii) fakes from commercial models (with identical bonafide data). Results (Table~\ref{tab:experiment3_language_system}) show that commercial models produce significantly harder fakes, yielding higher EERs. Further system-level analysis reveals that HeyGen samples are relatively easier to detect, while ChatGPT-4o-mini-TTS and ElevenLabs generate highly challenging fakes that push detection limits.  

\subsubsection{Language Analysis}


Finally, we evaluated language-specific performance using speech generated by commercial systems, since these represent the most realistic and challenging scenarios. Results (Table~\ref{tab:experiment3_language_system}) show clear variability: Vietnamese is the easiest to detect, whereas Tamil and Malay are the hardest, even for MoLEx\(_{ft}\). These findings highlight the influence of linguistic factors such as tonal vs. non-tonal properties on deepfake detectability, underscoring the need for language-aware countermeasures.
\begin{table}[htbp]
\centering
\def\arraystretch{1.15}
\setlength{\tabcolsep}{8.5pt}
\setlength{\abovetopsep}{0pt}
\setlength\belowbottomsep{0pt} 
\setlength\aboverulesep{0pt} 
\setlength\belowrulesep{0pt}
\caption{
Equal Error Rate (EER \%) of MoLEx before and after fine-tuning on the SEA-Bench benchmark.
\textbf{(1)} compares fake speech generated by open-source models (SEA Open-Sourced) 
against that generated by online commercial systems (SEA-Fake Online).  
\textbf{(2)} presents the breakdown of SEA Online by language, showing the cross-lingual robustness of MoLEx before and after fine-tuning.  
\textbf{(3)} presents the breakdown of SEA Online by system, comparing detection performance on speech generated by different online commercial TTS/VC providers.}
\label{tab:experiment3_language_system}
\vspace*{0.1in}
\begin{tabular}{lcc}
\toprule[1.0pt]
\textbf{Dataset} & \textbf{MoLEx} & \textbf{MoLEx-fine tuned} \\
\hline
\multicolumn{3}{l}{\textbf{(1) Open-sourced vs Commercial (Overall)}} \\
SEA Open-Sourced (b) & 35.80\% & 0.19\% \\
SEA commercial (c)     & 61.40\% & 0.28\% \\
\hline
\multicolumn{3}{l}{\textbf{(2) Commercial – Language-specific (d)}} \\
Hindi      & 64.50\% & 0.06\% \\
Tamil      & 68.60\% & 0.08\% \\
Malay      & 58.70\% & 1.35\% \\
Indonesian & 59.50\% & 0.07\% \\
Vietnamese & 48.50\% & 0.01\% \\
Thai       & 63.60\% & 0.11\% \\
\hline
\multicolumn{3}{l}{\textbf{(3) Commercial – System-specific (e)}} \\
MiniMax         & 56.50\% & 0.01\% \\
ChatGPT-4o-mini-TTS & 67.30\% & 0.08\% \\
Elevenlabs      & 65.50\% & 0.17\% \\
HeyGen          & 44.39\% & 1.35\% \\
\toprule[1.0pt]

\end{tabular}
\end{table}

\subsubsection{Key Takeaway}
SEA-Spoof exposes critical blind spots in existing benchmarks: models that excel on ASVspoof collapse when applied to SEA languages, highlighting the urgent need for regional coverage. At the same time, fine-tuning on SEA-Spoof dramatically restores detection accuracy, proving its value as both a diagnostic tool and a solution. By revealing where current models fail and demonstrating how those failures can be overcome, SEA-Spoof establishes a new foundation for building robust, cross-lingual, and fraud-resilient audio deepfake detection systems.

\section{Conclusion and Future Work}
\label{sec:conclusion}

We presented \textbf{SEA-Spoof}, the first large-scale dataset for audio deepfake detection in six South-East Asian languages. By pairing Bona-fide recordings with spoof speech generated from ten open-source and four commercial systems, SEA-Spoof provides a balanced and linguistically diverse benchmark that captures both real–fake variability and system-level heterogeneity. Our experiments reveal a stark reality: models that excel on established benchmarks fail when evaluated on SEA languages, showing poor transferability not only to commercial systems but also to open-source synthesis methods already covered in existing spoofing datasets. Fine-tuning on SEA-Spoof dramatically restores performance, demonstrating its dual role as a diagnostic benchmark for evaluating generalization gaps and as a practical resource for improving cross-lingual robustness.

For future work, we will extend SEA-Spoof to cover more dialects and lower-resource languages, while continuously integrating emerging synthesis technologies. We also plan to explore robust detection strategies, such as cross-source domain adaptation, language-aware modeling, and adversarially trained countermeasures, to strengthen the reliability of audio deepfake detection in practical scenarios. By filling a critical regional and linguistic gap, SEA-Spoof lays the foundation for developing truly robust, multilingual, and fraud-resilient deepfake detection systems.


\bibliographystyle{IEEEbib}
\bibliography{strings,refs}

\end{document}